\documentclass[%
 reprint,
 amsmath,amssymb,
 aps,
]{revtex4-1}

\usepackage{multirow}
\usepackage{bm}
\usepackage{graphicx}
\usepackage{braket}
\usepackage{physics}
\usepackage{amsmath}
\usepackage[colorlinks=true, allcolors=blue]{hyperref}
\usepackage{jabbrv}

\begin{document}

\preprint{APS/123-QED}

\title{Quantum entanglement patterns in the structure of atomic nuclei within the nuclear shell model}

\author{A. P\'{e}rez-Obiol}
\email{axel.perezobiol@bsc.es}
\affiliation{Barcelona Supercomputing Center, 08034 Barcelona, Spain}

\author{S. Masot-Llima}
\email{sergi.masot@bsc.es}
\affiliation{Barcelona Supercomputing Center, 08034 Barcelona, Spain}

\author{A. M. Romero}
\email{a.marquez.romero@fqa.ub.edu}
\author{J. Men\'{e}ndez}
\email{menendez@fqa.ub.edu}
\author{A. Rios}
\email{arnau.rios@fqa.ub.edu}
\affiliation{Departament de Física Quàntica i Astrofísica (FQA), Universitat de Barcelona (UB), c. Martí i Franqués, 1, 08028, Barcelona, Spain\\
Institut de Ciències del Cosmos (ICCUB), Universitat de Barcelona (UB), c. Martí i Franqués, 1, 08028 Barcelona, Spain}

\author{A. Garc\'{i}a-S\'{a}ez}
\email{artur.garcia@bsc.es}
\affiliation{Barcelona Supercomputing Center, 08034 Barcelona, Spain\\
Qilimanjaro Quantum Tech, 08007 Barcelona, Spain}

\author{B. Juli\'{a}-D\'{i}az}
\email{bruno@fqa.ub.edu}
\affiliation{Departament de Física Quàntica i Astrofísica (FQA), Universitat de Barcelona (UB), c. Martí i Franqués, 1, 08028 Barcelona, Spain\\
Institut de Ciències del Cosmos (ICCUB), Universitat de Barcelona (UB), c. Martí i Franqués, 1, 08028 Barcelona, Spain}

\date{\today}

\begin{abstract}
Quantum entanglement offers a unique perspective into the underlying structure of strongly-correlated systems such as atomic nuclei.
In this paper, we use quantum information tools to analyze the  structure of light and medium-mass berillyum, oxygen, neon and calcium isotopes within the nuclear shell model.
We use different entanglement metrics, including single-orbital entanglement, mutual information, and von Neumann entropies for different equipartitions of the shell-model valence space
and identify mode-entanglement patterns related to the energy, angular momentum and isospin of the nuclear single-particle orbitals.
We observe that the single-orbital entanglement is directly related to the number of valence nucleons and the energy structure of the shell,
while the mutual information 
highlights signatures of proton-proton and neutron-neutron pairing, as well as nuclear deformation.
Proton and neutron orbitals are weakly entangled by all measures, and in fact have the lowest von Neumann entropies
among all possible equipartitions of the valence space. In contrast, orbitals with opposite angular momentum
projection have relatively large entropies, especially in spherical nuclei.
This analysis provides a guide for designing more efficient quantum algorithms 
for the noisy intermediate-scale quantum era.
\end{abstract}

\maketitle

\section{Introduction}\label{sec1}

Entanglement is a fundamental concept in quantum mechanics~\cite{horodecki}. It characterizes correlations between particles or, in general, partitions within a system that can not be described independently of one another.   
Quantum many-body systems also show signatures of entanglement, with specific features in many-fermion systems~\cite{yu2003,Banuls2007,Gigena2015}. In addition, quantum entanglement is important from a theoretical point of view.
Entanglement properties typically undergo significant changes
near phase transitions, such as in spin and Fermi-Hubbard \cite{gu2004} systems at their critical point.
In high-energy physics, maximal entanglement has been used to constrain the coupling structure
of QED \cite{cervera2017}. In contrast, entanglement suppression
has been conjectured to be a property of low-energy strong interactions~\cite{beane}.

Quantum or classical simulations of many-body systems may be hampered if the entanglement structures couple different partitions. 
A sound understanding of the entanglement features of quantum many-body systems may thus be key to more efficient simulations.
Consider, for instance, a single partition of a given fermionic system. 
Low entanglement between two parts of a system may allow for simpler simulations for each
of the subsystems. If these simulations can be complemented with an effective way to integrate the residual entanglement between the partitions, such strategy may lead to results with a minor loss in precision at a fraction of the computational cost.
Analogously, ground states in condensed matter systems typically follow an area law, meaning that entanglement scales with the boundary of the partition, rather than with its volume~\cite{eisert2010colloquium}.
This allows one to use techniques such as density matrix renormalization group~\cite{white1992density} or tensor networks~\cite{orus2014practical,perezobiol_2022} to efficiently simulate large systems.

\begin{figure*}[t]
     \centering
     \includegraphics[width=0.75\linewidth]{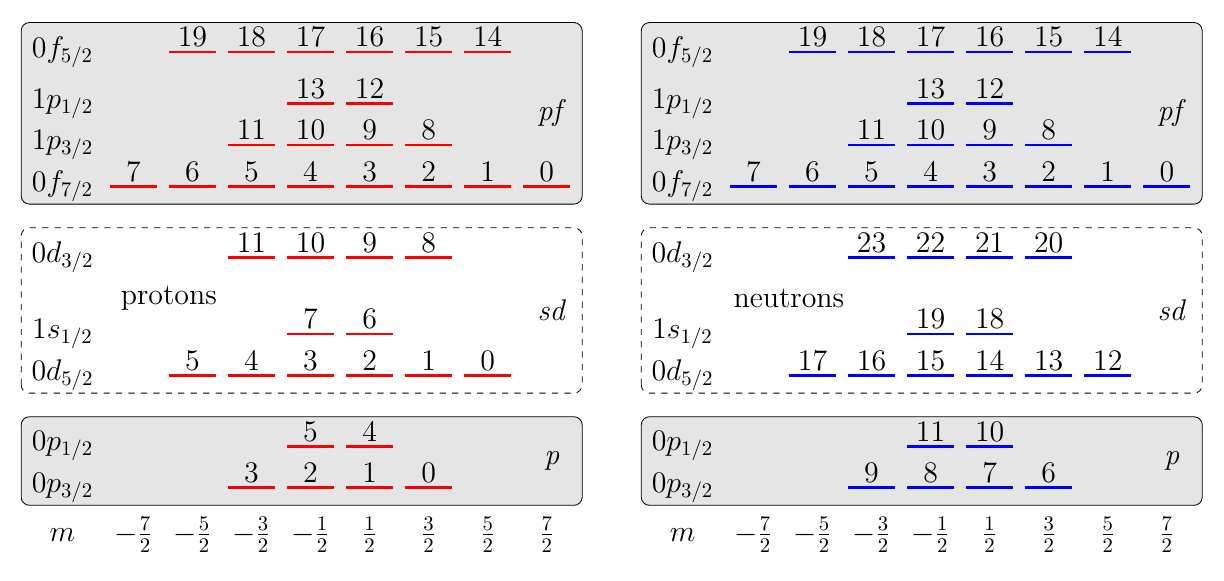}
     \caption{Diagram illustrating the valence spaces employed to describe all nuclei considered in this work. Single-particle states, shown in parallel if they correspond to the same $nl_j$ subshell, are labelled according to a number shown on top of them, for convenience. Combinations of protons and neutrons occupying these single-particle states form the many-body basis of the nucleus. For nuclei described in the \emph{sd} shell, the \emph{p} shell is assumed to be a fully occupied inert core, and the \emph{pf} shell is assumed to be completely empty. Excitations to states outside the valence space, marked in grey, are not allowed.
     } %
     \label{fig:sd_diagram}     
\end{figure*}

However, in nuclear physics, entanglement has been much less studied. This is
in part due to the complex nature of the nuclear force, but also due to the difficulty
to relate entanglement to measurable observables.
In very light nuclei, like ${}^4$He and ${}^6$He, the entanglement structure was 
found to be highly dependent on the many-body basis~\cite{Robin2021}.
In nucleon-nucleon scattering, proton-neutron pairs are found to be entangled~\cite{bai2023spin,miller2023entanglement}. 
Yet, nuclear shell model simulations of mid-mass isotopes indicate that protons and neutrons show very little entanglement~\cite{johnson2023proton}. 
Entanglement is also relevant for the dynamics of nuclear reactions~\cite{Bulgac:2022ygo}.
Interestingly, a proposal has been recently put forward, indicating that nuclear matter follows a volume law, instead of an area law~\cite{gu2023entanglement}.

Entanglement plays a pivotal role in quantum-resource quantification for quantum computation, communication, and sensing.
In the context of quantum simulations, variational algorithms have
been devised and tested to reproduce ground states of quantum
many-body systems. In particular, Ref.~\cite{stetcu} analyzed single- and double-orbital entanglement for ${}^8$Be within the nuclear shell model. Entanglement measures were found to be
almost maximal, as a consequence of the strong correlations in
the the \emph{p} shell.  In addition, Ref.~\cite{paulin} explored entanglement-based partitions in this configuration space using 
neural networks for optimal simulations. In heavier systems, entanglement was first studied within the nuclear shell model using density matrix renormalization group (DMRG) tools~\cite{Legeza:2015fja}, followed by recent studies on two-nucleon configurations~\cite{Kruppa:2020rfa} and the seniority model~\cite{Kruppa:2021yqs}. More broadly, single-orbital entanglement within the nuclear shell model has been studied with an adaptive variational quantum eigensolver~\cite{grimsley2019adaptive} for various nuclei across the \emph{p}, \emph{sd} and \emph{pf} shells~\cite{PerezMarquez2023}. Very recently, Tichai et al.~\cite{tichai2022combining} used various entanglement measures to study in detail the structure of $sd-$shell and nickel isotopes within the \emph{ab initio} valence-space in-medium similarity renormalization group. Further, they combined this approach with the DMRG to optimize the convergence of the many-body calculations.

In this work, we complement these previous studies by systematically analyzing mode entanglement in nuclear shell model ground states.
We provide an overall picture of the entanglement structure for the isotopic chains of Be, O, Ne and Ca, studied in the \emph{p}, \emph{sd}, and \emph{pf} configuration spaces. 
Figure~\ref{fig:sd_diagram} shows a diagram representing the corresponding configuration spaces. 
We use three different entanglement measures: single-orbital entanglement, mutual information, and von Neumann entropies, for the most natural equipartitions of 
these configuration spaces, 
providing compelling insights into the entanglement of the different divisions.

This article is organized as follows: in Section~\ref{sec:nsm} we provide an introductory outline of the nuclear shell model, our method of choice to find ground states of nuclei. In Section~\ref{sec:ent}, we introduce the different measures that we use to quantify the entanglement in the nuclear configuration space. Finally, we present our results in Section~\ref{sec:results} and our conclusions in Section~\ref{sec:conclusions}.

\section{Overview of  the nuclear shell model}
\label{sec:nsm}

The nuclear shell model is one of the most successful theories of nuclear structure~\cite{heyde1994nuclear,de2013nuclear}. It considers nuclei as composite systems of protons and neutrons, or nucleons, that interact with each other in a restricted configuration space, customarily called valence space. The nuclear interaction is rotationally invariant, and it is usually considered to be symmetric under proton-neutron exchange. One of the main features of the nuclear interaction is a spin-orbit term responsible for the so-called \emph{magic numbers}: special combinations of protons ($Z$) and neutrons ($N$) building up particularly stable, spherical nuclei~\cite{mayerII,Haxel:1949fjd}. This justifies the main assumption of the shell model, that nuclear dynamics can be approximated by the many-body configurations built in a valence space limited by two magic numbers. The valence spaces considered in this work are presented in Fig.~\ref{fig:sd_diagram}. Single-particle states below the valence space are fully occupied and form an inert core, whereas states above are truncated based on the large energy gaps between magic number configurations.

Given the symmetries of the nuclear interaction between particles in the valence space, the single-particle states (or single-particle orbitals) can be labelled using a set of quantum numbers $\{n,l,j,m,t_z\}$. These correspond to the principal quantum number $n$, the orbital angular momentum $l$, the total angular momentum $j$ (resulting from the coupling of $l$ with the spin $s=1/2$ of nucleons) and its third-component projection $m$. The third-component projection of the isospin, $t_z$, specifies if a nucleon is a proton or a neutron. The corresponding $2j+1$ energy-degenerate single-particle states are grouped into the $nl_j$ subshells, as shown in Fig.~\ref{fig:sd_diagram}.

In a second quantization scheme, the effective Hamiltonian in the valence space reads
\begin{equation}\label{eq:smham}
    H_{\rm{eff}} = \sum_i \varepsilon_i a_i^{\dag} a_i + \frac{1}{4} \sum_{ijkl}\bar{v}_{ijkl}
    a_i^{\dag} a_j^{\dag} a_l a_k,
\end{equation}
where $\varepsilon_i$ is the energy of the single-particle state $i$, $\bar{v}_{ijkl} = v_{ijkl} - v_{ijlk}$ are antisymmetrized two-body matrix elements and $a_i$ ($a_i^{\dag}$) are particle annihilation (creation) operators associated to the state $i$. In this work, we use standard phenomenological Hamiltonians, with components adjusted to reproduce key properties of selected nuclei~\cite{Poves:1981zz}. These Hamiltonians describe very well the low-energy properties of light and medium-mass nuclei across the nuclear chart~\cite{brown1988status,Caurier:2004gf,otsuka2020evolution}. Effective Hamiltonians can also be derived based on an effective theory of the fundamental theory of the nuclear force, quantum chromodymamics, using so-called \emph{ab initio} techniques~\cite{Stroberg:2019mxo,dikmen,jansen2014,hebeler2015} like in Ref.~\cite{tichai2022combining}.

Many-body states in the valence space are described employing antisymmetrized products of single-particle states, also referred to as Slater determinants. A standard choice to build this many-body basis is to use the \emph{M-}scheme, in which Slater determinants have a well-defined third component $M$ of the total angular momentum $J$. Because of the properties of the $SU(2)$ algebra of angular momentum~\cite{varshalovich1988quantum}, $M$ is simply the sum of the total 
$m$ components of the single-particle states occupied by the nucleons. These many-body states form a basis of the valence space, and the ground and excited states of the nucleus can be expanded as
\begin{equation}\label{eq:stateJT}
   | JM\,TT_z\rangle = \sum_{\alpha} c_{\alpha}  | \alpha, M T_z\rangle,
\end{equation} 
where the $c_{\alpha}$ coefficients are obtained by solving the many-body Schr\"odinger equation, for instance through the diagonalization of the Hamiltonian matrix in the many-body basis~\cite{caurier1999antoine,brown2014shell,johnson2018bigstick,Shimizu:2019xcd}. These eigenstates have good angular momentum $J$ and isospin $T$ 
quantum numbers, with corresponding third-component projections $M$ and $T_z$. State-of-the-art shell-model codes use sophisticated Lanczos methods for the Hamiltonian diagonalization, which often require classical supercomputuing resources. 

The nuclear shell model is a reference method for light- and medium-mass nuclei, but calculations become unattainable for heavy nuclei. As the number of valence nucleons increases, the number of many-body states in the valence space grows exponentially, quickly reaching a bottleneck where calculations are no further feasible with current classical supercomputers. Quantum information tools may help identify crucial correlations in the shell-model valence space and may facilitate systematic and well-controlled truncation protocols, that include the most relevant degrees of freedom~\cite{momme2023}. While this challenge is important from a fundamental nuclear structure point of view, it is also pertinent to optimise the performance of quantum simulations in the noisy intermediate-scale quantum era. Promising implementations of the nuclear shell model in digital quantum computers have been
proposed using variational quantum eigensolvers~\cite{papenbrock,romeroquantum} and quantum Lanczos~\cite{kirby2023exact} algorithms.

\section{Entropy and mutual information for entanglement assessment}
\label{sec:ent}
Entanglement quantifies the inseparability between quantum systems. When two systems $A$ and $B$, characterized by states $\ket{\psi_A}$ and $\ket{\psi_B}$, are entangled,
the complete state $\ket{\psi}$ can not be written in the form 
\begin{equation}\label{eq:separable_state}
    \ket{\psi} = \ket{\psi_A} \otimes \ket{\psi_B}, 
\end{equation}
that is, as a tensor product of the individual states. If the states are separable, as opposed to entangled, the statistics and behavior of each system can be treated independently. 
This is why, when there is low entanglement, classical resources can simulate quantum systems efficiently (although this is not the only case \cite{scott2004}). Therefore, finding partitions that exhibit low entanglement is of utmost importance.

In fact, 
a natural split is already present in the nuclear shell model. As discussed in Sec.~\ref{sec:nsm}, 
the \emph{a priori} separation between the inert core, the valence space and the excluded space assumes that there is no entanglement between these spaces.
In this work, we focus on entanglement measures in the valence space, where additional insight of the entanglement structure could improve nuclear shell-model calculations. 
We specifically consider bipartite entanglement, considering two generic partitions 
within the system.
Although multipartite entanglement is complex and still a subject of intensive study~\cite{walter2016multipartite}, the case of bipartite entanglement is well understood~\cite{horodecki} and linked to quantitative metrics usually referred to as entropies. 

In this context, a standard choice is the von Neumann entropy $S$, defined as 
\begin{equation}\label{eq:vnentropy}
    S(\rho) = - \text{Tr}(\rho \log_2 \rho) = - \sum_i \rho_i \log_2 \rho_i,
\end{equation}
where $\rho_i$ are the eigenvalues of the density matrix $\rho$, and where the logarithm basis is $2$ for qubits. For a state $\ket{\psi}$, the density matrix is pure, $\rho=\ket{\psi}\bra{\psi}$. Pure density matrices have a single non-zero eigenvalue, and consequently no von Neumann entropy, $S(\rho)=0$. When considering a partition of the whole system into subsystems $A$ and $B$, the reduced density matrix of subsystem $A$ is obtained by tracing out the degrees of freedom of subsystem $B$ from $\rho$, that is $\rho_A = \Tr_B(\rho)$. If the state of the whole system is separable, as in Eq.~(\ref{eq:separable_state}), the corresponding trace results in a pure state
\begin{equation}
    \rho = \ket{\psi_A} \otimes \ket{\psi_B} \bra{\psi_A} \otimes \bra{\psi_B} \rightarrow \rho_A = \ket{\psi_A}\bra{\psi_A} .
\end{equation}
As a consequence, the von Neumann entropies of the subsystems are also zero, 
$S(\rho_A)\equiv S(A) = 0$. In contrast, for a Bell-type entangled state $\ket{\psi} = (\ket{0}_A\ket{0}_B + \ket{1}_A\ket{1}_B)/\sqrt{2}$, 
we find
\begin{equation}
    \rho_A= \frac{1}{2} \ket{0}\bra{0} + \frac{1}{2} \ket{1}\bra{1} \rightarrow S(A) = 1 \, .
\end{equation}
This highlights how the von Neumann entropy quantifies our notions of entanglement for bipartitions.

\begin{figure}[t]
    \centering
    \includegraphics[width=1.\linewidth]{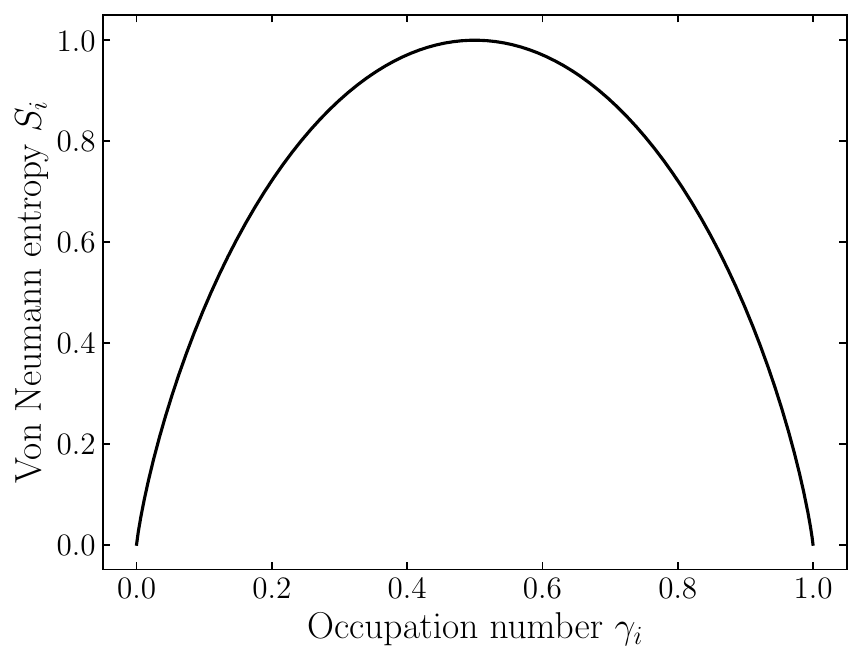}
    \caption{Single-orbital entropy $S_i$ from Eq.~(\ref{eq:sp_entropy}) as a function of the occupation probability $\gamma_i$ of the single-particle state $i$. The state is maximally entangled when the occupation probability is 50\%.}
    \label{fig:sp_entropy}
\end{figure}

An illustrative example is the partition of one single-particle orbital and the rest of the system. In this case, the entropy has already been linked to the occupation number~\cite{yu2003,Gigena2015} and, more recently, to the choice of single-particle basis in the shell model~\cite{Robin2021}. Using the Jordan-Wigner fermionic mapping~\cite{jw}, where a qubit corresponds to a single-particle orbital being empty or occupied, we can use Eq.~(\ref{eq:vnentropy}) for the single-particle bipartition of particle-number conserved states to show that the
entropy is
\begin{equation}\label{eq:sp_entropy}
    S_i = -\gamma_i \log_2\gamma_i - (1-\gamma_i)\log_2(1-\gamma_i),
\end{equation}
where $\gamma_i = \langle \psi |a_i^{\dag} a_i| \psi \rangle$ corresponds to the occupation number (or occupation probability) of the single-particle orbital $i$ in a system described by $|\psi\rangle$.
$S_i$ can be simplified to the expression above because, for a conserved number of particles in the statevector,
the non-diagonal elements in the one-qubit reduced density matrix used as input in Eq.~(\ref{eq:vnentropy}), $|0\rangle\langle1|$ and $|1\rangle\langle0|$, vanish,
while the diagonal ones give the probabilities of finding that qubit in state 0 or 1.
Figure~\ref{fig:sp_entropy} shows the single-orbital entropy $S_i$ as a function of the occupation number $\gamma_i$. 
Single-orbital entropies are maximal when the occupation numbers are as likely to be filled than to be empty, $\gamma_i=1/2$. 
In contrast, states that are almost fully occupied or fully unoccupied have near zero single-orbital entropy.

In all our calculations of entanglement we use Eq.~(\ref{eq:vnentropy}), obtaining the reduced density matrix through the partial trace of one of the partitions. Thus, we do not explicitly compute orbital occupation numbers, which is not always possible.

For example, partitioning the system into two qubits and tracing out the remaining degrees of freedom results in a reduced density matrix that is in general not diagonal.
In this case, even for states conserving the number of particles,  
the reduced density matrix contains finite non-diagonal terms $|01\rangle\langle10|$, $|10\rangle\langle01|$~\cite{Boguslawski2015}. However, if the two qubits correspond to a proton and a neutron orbital,
or orbitals with different angular momentum projection $m$,
the non-diagonal terms vanish due to isospin and angular momentum conservation. The entropy can be directly evaluated using Eq.~(\ref{eq:vnentropy}) again,
where now the diagonal entries of the reduced density matrix correspond to the probabilities for qubits $i$ and $j$ being empty or occupied, $\langle \psi|(1-n_i)(1-n_j)|\psi\rangle$, $\langle \psi|(1-n_i)n_j|\psi\rangle$,
$\langle \psi|n_i(1-n_j)|\psi\rangle$, and $\langle \psi|n_i n_j|\psi\rangle$,
with $n_i=a_i^\dag a_i$.
For this particular case, the general Eq.~(\ref{eq:vnentropy}) can thus be simplified in terms of orbital occupation numbers.

We also take into consideration other useful entanglement metrics. The conditional entropy, $S(A|B)$, illuminates the dependence of subsystem $A$'s degrees of freedom on subsystem $B$. It is calculated as 
\begin{equation}\label{eq:cond_entropy}
    S(A|B) = S(AB) - S(B),
\end{equation}
where $S(AB)$ is the entropy of the joint system. Using this metric, we can describe the mutual information of two systems $A$ and $B$,
\begin{equation}\label{eq:mutual_info}
\begin{split}
        S(A;B) & = S(A) - S(A|B) \\ & = S(A) + S(B) - S(AB),
\end{split}
\end{equation}
which we use in this work with the simplified notation $S_{A,B}$. In particular, we define the mutual information between two orbitals $i$, $j$ as $S_{ij}\equiv\frac{1}{2}S(i;j)$, where the factor $\frac{1}{2}$ is included such that it ranges between 0 and 1. The mutual information is symmetric under the exchange of its arguments. It provides an insight on how much subsystems $A$ and $B$ are correlated when ignoring the degrees of freedom of the rest of the system. Specifically, in the context of $A$ and $B$ being subsystems of a larger system $ABC$, a low amount of mutual information unveils that, even if the state of the total system is not separable in states of subsystems $A$ and $B$ (and thus entangled), such entanglement is linked to $C$ and not contained in $AB$. 
\subsection{Fermionic systems}

In the case of fermionic systems,
quantifying entanglement is especially challenging because we lack a well-defined underlying separable space, as used in the definition of Eq.~(\ref{eq:separable_state}). Fermions are identical particles which fulfil Pauli's exclusion principle, and a many-body fermionic state must be antisymmetric.
For example, two spin$-1/2$ fermions can couple their spins to form a singlet state, with total spin $0$, in first quantization.
In second quantization, on the other hand, this corresponds to creating two modes on the vacuum state, $a_\uparrow^\dagger a_\downarrow^\dagger \ket{0},$ using mode creation operators of spin projections up and down. With the right encoding of fermions onto qubits, this ends up as a separable state $\ket{11}$, 
\begin{equation}\label{eq:singlet}
    \underbrace{\frac{1}{\sqrt{2}} \big( \ket{\uparrow \downarrow} - \ket{\downarrow \uparrow} \big)}_\text{singlet state} 
    =
     a_\uparrow^\dagger a_\downarrow^\dagger \ket{0}
    ~\xrightarrow{\text{encoding}}~
    \ket{11}.
\end{equation}
This separable expression is in contrast with the first-quantization expression.
In other words, the singlet state in the first-quantized particle basis
is considered maximally entangled, while the corresponding qubit state
has zero entanglement.

We note that these anomalies are directly related to the indistinguishability of particles. In other words, any partition that separates distinguishable particles, such as neutrons and protons, does not exhibit this problem. Entanglement quantification measures of bipartitions of identical particles, however, need to address this issue. 
Different approaches to a proper characterization have been proposed~\cite{Benatti2020,Johann2021}, favouring those in second quantization for their consistency. This motivates our choice of encoding.
 
In the implementation of fermionic systems on quantum circuits, the encoding between qubits and single-particle degrees of freedom is very important. The advantages of different fermionic mappings have been studied extensively~\cite{bravyi-kitaev2012,setia2019,whitfield2016}, although usually under the scope of performance and scalability. In other words, the focus has been on 
how efficiently one can encode a specific system in terms of number of qubits and circuit depth.  Because operators on different qubits commute freely, but operators on fermions do not, each encoding must balance the locality of the original system's degrees of freedom against a method to modify the system's parity each time one acts on the state. One of the most common fermionic mappings, the Jordan-Wigner encoding~\cite{whitfield2011}, lies at one extreme. Qubits correspond exactly to single-particle states in the fermionic system, at the cost of local operators on the fermions becoming completely delocalized on the qubits. 

In our analysis of entanglement in the nuclear shell model, we use the Jordan-Wigner encoding because it becomes advantageous on two fronts. Firstly, it allows us to simplify the treatment of fermionic entanglement by using second quantization. Having a fixed particle number avoids the need for more complex figures of merit~\cite{Gigena2015}. Secondly, it provides a direct connection between specific qubits and single-particle orbitals, as indicated by the labels in Fig.~\ref{fig:sd_diagram}.
In actual circuit simulations, the Jordan-Wigner encoding may increase the circuit depth compared to other encodings, due to the non-locality of the encoded fermionic operators. This disadvantage, however, may be offset if one finds 
low-entangled partitions, which allow for more efficient simulations.

\subsection{Maximal entropy states}\label{sec:max_entropy}

Entanglement measures can only be used to identify relevant quantum features if there is a notion of maximal entanglement. By construction, the nuclear shell model constrains the number of allowed many-body states to those included in the valence space, and therefore the maximal entropy. This means that it is not sufficient to focus on the dimension of the Hilbert space obtained after the fermionic encoding.
For example, let us consider $^{8}$Be, with two valence neutrons and protons in the \emph{p} shell (see Fig.~\ref{fig:sd_diagram}). The $^{8}$Be ground state has $J=M=0$. Since there are $12$ single-particle states, we need $12$ qubits in the Jordan-Wigner mapping to 
encode all the possible fermionic excitations. The dimension of the Hilbert space in this computational basis is thus $2^{12}$. After an arbitrary equipartition, the resulting spaces of the subsystems would have a dimension of $2^6$. However, to reach the maximal entropy $S_{max}$ between two partitions $A$ and $B$, one must actually be able to build a state of the form
\begin{equation}\label{eq:max_ent}
    \ket{\Psi}_{shell} = \sum_i^{2^{S_{max}}} c_i \ket{\psi_i}_A \otimes \ket{\phi_i}_B ,
\end{equation}
according to the Schmidt decomposition \cite{ekert_1995}. This, however, is not possible for $S_{max}=6$ in $^{8}$Be, due to proton and neutron number conservation and the constraint $M=0$. 
Therefore, the dimension of this truncated space after a partition is only an upper bound for the von Neumann entropy, and is in fact unreachable. A better bound would be the dimension of the many-body basis when considering the modes in the partition.
We can  find all possible product states by running through each possible $M_Z\in \{ -2,-1,0,1,2\}$ value for protons, and pairing them with any of the neutron states with opposite $M_N=-M_Z$. In the proton-neutron partition, there are $15$ possible ways to arrange the $2$ protons in $^{8}$Be, and we can pair each of these with the neutron state that mirrors the occupations over the sign of $M$ to form a state of the form in Eq.~(\ref{eq:max_ent}). Constructing a state with higher entropy is not possible because there are no more elements of the basis, so $S=\log_2(15)=3.9<6$ is the maximum limit.
An additional caveat must be considered when looking at general bipartitions. Let us consider the proton-neutron partition again, in a nucleus with more neutrons than protons below the half-filling of the valence space, as for example $^{10}$Be or $^{22}$Ne. Clearly, the neutron many-body basis has a bigger dimension and conditions which of the two partitions limits the entropy, $S$. For general bipartitions, however, there is no guarantee that the smallest dimension of the two bipartitions limits the maximum entropy. Let us illustrate this with the bipartition of orbitals with opposite $m$ for $^{8}$Be, which seems to maintain the symmetry across protons and neutrons. We can find some combinations where two states on the $m>0$ partition ($[1,7,10]$ and $[4,7,10]$ following the numbering of the $p$ shell in Fig.~\ref{fig:sd_diagram}) that can only be paired with one basis element in the other one ($[3]$), meaning that they can not contribute to $S$ as two separate elements of Eq.~(\ref{eq:max_ent}).
In principle, straightforward algorithmic efforts to check all combinations for a given bipartition are unreachable by classical computation due to their exponential scaling. A more naive approach is doing the pairing only for a few examples. Even if this is non-exhaustive, it already provides a lower bound for the maximal entropy at a small computational cost. This is enough to decide whether the entropy of that partition is small in relative terms - it can only become smaller by saturating the bound - and we can do so for all partitions. 
In addition, we systematically check that for the most characteristic proton-neutron partition these bounds are actually satisfied.
In the following section, we highlight two characteristic partitions based on physical intuition. First, we look into a proton-neutron partition. Moreover, we also discuss bipartitions formed by states with opposite values of $m$. In both cases, we can compute the corresponding entropy bounds and compare whether shell-model simulations are close to saturating them.

\section{Results}
\label{sec:results}

\begin{table}[t]
\begin{center}
\begin{tabular}{c|c|c}
   \text{Nucleus} & $S_i$ \text{(proton shells)} & $S_i$ \text{(neutron shells)}  \\ \hline
  $^8{}$Be    &  0.95, 0.85 &  0.95, 0.85 \\
  $^{10}{}$Be &  0.98, 0.61 &  0.66, 0.92 \\
  $^{12}{}$Be &  0.99, 0.54 &  - \\\hline
  $^{18}{}$O  &  -           &  0.82, 0.67, 0.17  \\
  $^{20}{}$O  &  -           &  0.98, 0.66, 0.30 \\
  $^{22}{}$O  &  -           &  0.45, 0.68, 0.29 \\
  $^{24}{}$O  &  -           &  0.18, 0.21, 0.31 \\
  $^{26}{}$O  & -            &  0.11, 0.14, 1.00 \\ \hline
  $^{20}{}$Ne & 0.73, 0.80, 0.36  &  0.73, 0.80, 0.36 \\
  $^{22}{}$Ne & 0.80, 0.71, 0.24  &  1.00, 0.71, 0.50 \\
  $^{24}{}$Ne & 0.86, 0.38, 0.21  &  0.64, 0.84, 0.50 \\
  $^{26}{}$Ne & 0.85, 0.50, 0.20  &  0.30, 0.63, 0.63 \\
  $^{28}{}$Ne & 0.88, 0.31, 0.15  &  0.14, 0.23, 0.99 \\ \hline
  $^{42}{}$Ca & -            &  0.78, 0.12, 0.07, 0.10 \\
  $^{44}{}$Ca & -            &  1.00, 0.16, 0.10, 0.15  \\
  $^{46}{}$Ca & -            &  0.86, 0.16, 0.11, 0.17 \\
  $^{48}{}$Ca & -            &  0.18, 0.13, 0.10, 0.14 \\
  $^{50}{}$Ca & -            &  0.18, 1.00, 0.44, 0.20 
\end{tabular}
\end{center}
\caption{Single-orbital entropies $S_i$
for Be isotopes in the $p$ shell, O and Ne nuclei in the $sd$ shell and Ca isotopes in the $pf$ shell. The entropies are equal for the $2j+1$ single-particle orbitals in the $nl_j$ subshells and they are shown in energy order from left to right: $0p_{3/2}$ and $0p_{1/2}$ in the $p$ shell; $0d_{5/2}$, $1s_{1/2}$ and $0d_{3/2}$ in the $sd$ shell and $0f_{7/2}$, $1p_{3/2}$, $1p_{1/2}$ and $0f_{5/2}$ in the $pf$ shell.  
Empty cells correspond to isotopes with either an empty-proton shell (as for O and Ca)
or a full neutron shell ($^{12}$Be), which trivially have $S_i=0$.
}
\label{tab:Si}
\end{table}

\begin{figure*}[t]
     \centering
     \includegraphics[width=\linewidth]{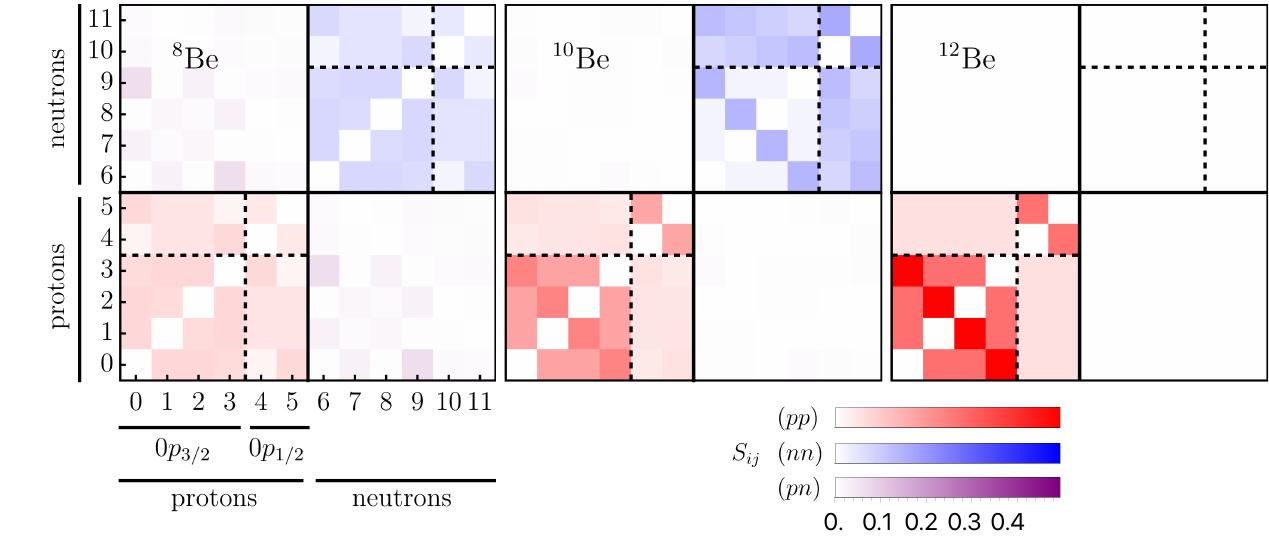}
     \caption{Mutual information $S_{ij}$ for $^{8,10,12}$Be in the $p$ shell. 
     Solid black lines divide the proton-proton (bottom-left),
     proton-neutron (top-left), neutron-proton (bottom-right), and neutron-neutron (top-right) sectors.
     The dashed lines  correspond to the two different subshells.
      Single-particle orbitals are sorted according Fig.~\ref{fig:sd_diagram}, as explicitly shown for $^8\text{Be}$ and reflected on the corresponding numbering of the axes.
     All mutual information matrices are symmetric, as requested by Eq.~(\ref{eq:mutual_info}). Elements in the diagonal, corresponding to single-particle entropies $S_i$, are arbitrarily set to $0$ (white) to showcase the inter-orbital behaviour.
    }
     \label{fig:SijBe}     
\end{figure*}
We study the entanglement properties of selected beryllium, oxygen, neon and calcium nuclei, all of which have
an even number of nucleons. The ground states of all
these nuclei have $J=M=0$, as defined in Eq.~(\ref{eq:stateJT}). We use this symmetry to build the many-body basis, including only Slater determinants with $M=0$.
We employ the Cohen-Kurath interaction in the $p$ shell~\cite{cohen1965effective}, USDB in the $sd$ shell~\cite{Brown2006} and KB3G in the $pf$ shell~\cite{Poves2001}.

\subsection{Single-orbital entanglement}

We start our discussion quantifying the single-particle
entanglement in different isotopes. 
The single-orbital entanglement entropy $S_i$, defined in Eq.~(\ref{eq:sp_entropy}), is a direct reflection of the single-orbital occupation number. In turn, this is intertwined with the subshell energy structure and the number of valence nucleons.
The entanglement between two sets of modes depends in the first place on the single-orbital entanglement, which becomes the dominant factor whenever there is a large energy difference between subshells, the so-called subshell closures. For instance, in the $pf$ shell, $N=28$ is a magic number. As discussed above, completely occupied or empty states are directly linked to near-zero single-particles entropies.

Consequently, isotopes with a number of neutrons equal to the number of orbitals in the lowest-energy subshells may have less entanglement entropy, while those with half-filled subshells have much more potential to be entangled. Let us provide an illustrative example using Ca isotopes. 
In $^{44}$Ca, with 4 valence neutrons in the $pf$ shell, the lowest and degenerate orbitals of the $0f_{7/2}$ subshell have an occupation number $\gamma_{0f_{7/2}}=0.477$ and, in consequence, almost maximal single-orbital entanglement $S_{0f_{7/2}}=0.998$. In contrast, the modes in the remaining subshells,
$1p_{3/2}$, $1p_{1/2}$ and $0f_{5/2}$,
are mostly empty, with occupations $\gamma_{1p_{3/2}}=0.023$, $\gamma_{1p_{1/2}}=0.013$, $\gamma_{0f_{5/2}}=0.022$. All these modes have low single-particle entropies, $S_i<0.2$. 
These entanglement properties are in stark contrast to those of ${}^{50}$Ca, which has 6 more neutrons. Here, the orbital occupations for each subshell are $\gamma_{0f_{7/2}}=0.972$, $\gamma_{1p_{3/2}}=0.465$, $\gamma_{1p_{1/2}}=0.091$, and $\gamma_{0f_{7/2}}=0.031$. The $0f_{7/2}$ single-particle entropy is now substantially lower than in ${}^{44}$Ca,
with $S_{0f_{7/2}}=0.184$, whereas the $0p_{3/2}$ states are almost maximally entangled, with $S_{0p_{3/2}}=0.996$.
That is, the $1p_{3/2}$ subshell shows the largest entanglement, as expected from a naive filling of the $pf$ shell. Actually, $N=32$ is also a magic number in Ca~\cite{Huck85,Gade:2006dp,Wienholtz:2013nya}. A similar discussion for the single-orbital entanglement in  oxygen and nickel isotopes calculated with an ab initio valence-space framework has been given in Ref.~\cite{tichai2022combining}.

We observe similar patterns for all the nuclei studied in this work. Table~\ref{tab:Si} lists the corresponding single-particle entropies of different states for all isotopes. All the results point to maximal single-particle entropies appearing in mid-subshell isotopes. 

Interestingly, Table~\ref{tab:Si} also presents some variation in the proton single-orbital entropies of different isotopes of the same element. For instance, the $0p_{1/2}$ orbital in $^8$Be and the $1s_{1/2}$ orbital in $^{20-22}$Ne show relatively high entropy, reflecting a larger proton occupation number than other isotopes. In fact, in $^{20}$Ne the relative occupation of the $1s_{1/2}$ mode is higher than the one of the $0d_{5/2}$ orbital, which sits at lower energy. This is an indication of nuclear correlations being important for these nuclei~\cite{Legeza:2015fja}: the quadrupole-quadrupole interaction makes these systems deformed~\cite{Elliott1,Elliott2}, and the large quadrupole correlation energy competes with the naive filling of the different modes according to their single-particle energy~\cite{Zuker1995}

Single-orbital entropy sets a bound for how much orbitals in a particular subshell can contribute to multi-orbital entanglement. Let us stress that while $S_i$ provides a measure of how much an orbital is entangled with the rest of the modes, it does not specify with which part of the nucleus it is entangled, nor distinguishes between single-particle states in each subshell with different angular-momentum projections, $m$.

\subsection{Mutual information}
\label{sec:mutual}

\begin{figure*}[t]
     \centering
     \includegraphics[width=\linewidth]{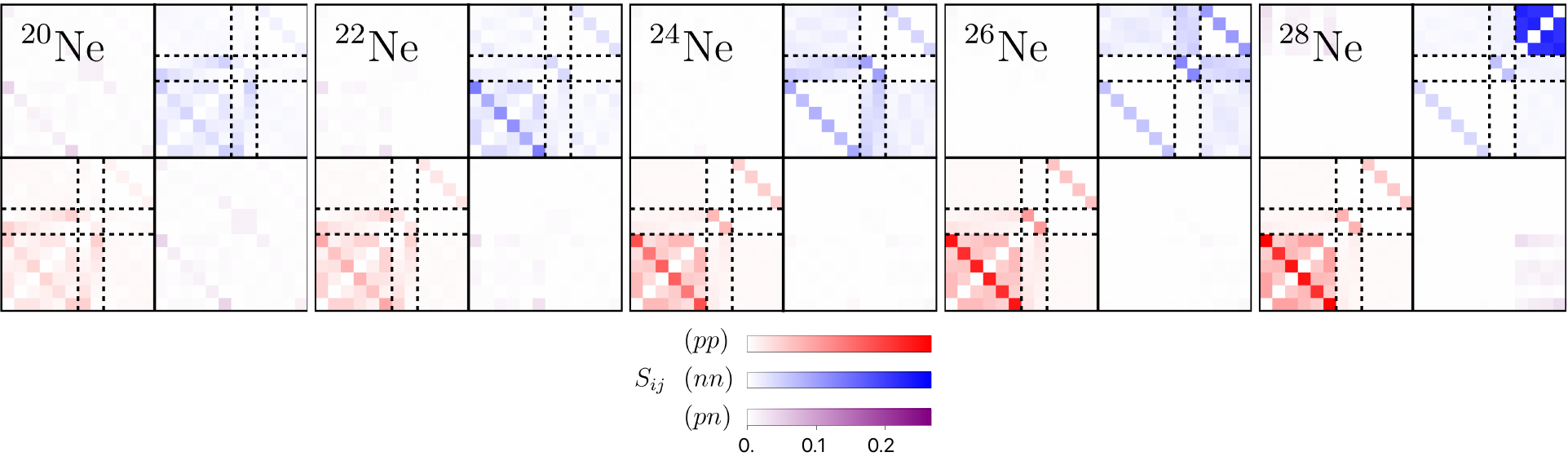}
     \caption{Mutual information $S_{ij}$ for $^{20-28}$Ne. 
     Numbering of orbitals and their organization follows the same scheme as in Fig.~\ref{fig:SijBe}, except now the dashed lines correspond, from left to right (and bottom to top), to the $0d_{5/2}$, $1s_{1/2}$ and $0d_{3/2}$ subshells of the $sd$ shell.
    }
     \label{fig:SijNe}     
\end{figure*}

A more general picture of the entanglement structure of the nucleus is given by the mutual information matrix, $S_{ij}$. Figure~\ref{fig:SijBe} shows the mutual information between all pairs of single-particle orbitals, $(i,j)$, for $^{8,10,12}$Be. The leftmost panel illustrates the structure of the mutual information matrix by explicitly labelling each orbital with the convention shown in Fig.~\ref{fig:sd_diagram}.
We organise the orbitals in neutron and proton blocks, with black solid lines 
separating proton-proton (bottom-left), proton-neutron (top-left), neutron-proton (bottom-right),
and neutron-neutron (top-right) correlations.
Proton-proton and neutron-neutron mutual information is colored in red and blue, respectively,
while the proton-neutron and neutron-proton sectors are shown in a purple colour scale.
The scale is the same for all isotopes and blocks within each isotope, with darker shades implying larger $S_{ij}$ values.
The subshell structure of each proton-proton and neutron-neutron block is illustrated by black dashed lines which separate subshells. In the $p$ shell, these correspond to the $0p_{3/2}$ and $0p_{1/2}$ subsells. 

Finally, within each subshell, the orbitals are sorted by the third component of angular momentum, $m$, following the notation of Fig.~\ref{fig:sd_diagram}.

The leftmost panel of Fig.~\ref{fig:SijBe} corresponds to ${}^8$Be, with 2 protons and 2 neutrons in
each of the 6-orbital valence spaces.
${}^8$Be shows a relatively low mutual information in all orbitals, although the like-particle mutual information is more prominent than the corresponding neutron-proton values. The central panel focuses on ${}^{10}$Be, which has the largest neutron-neutron entanglement among the three isotopes.
In the rightmost panel, for ${}^{12}$Be,  neutrons completely fill the valence space and proton-neutron, neutron-proton,
and neutron-neutron entanglement is trivially zero.
Proton-proton entanglement grows with the neutron excess, though,
and the proton-proton sector of ${}^{12}$Be shows the largest 
mutual information values of all three isotopes. In contrast, proton-neutron entanglement is relatively low or zero for all three isotopes in comparison with the like-particle entanglement.

The mutual information results for both ${}^{10}$Be and ${}^{12}$Be show a prominent feature that is shared by many of the other nuclei we study. Specifically, in agreement with previous works~\cite{Legeza:2015fja,tichai2022combining} we find that the mutual information is largest for orbitals with opposite angular-momentum projection $m$. These orbitals correspond to the diagonals in each subshell within the proton-proton and neutron-neutron sectors. We find that these diagonals are notably darker than the rest of the matrix, indicating larger entanglement among these specific partitions.

The patterns that we have identified so far, namely the relation of $S_{ij}$ with occupation numbers;  the relatively large mutual information  among orbitals with opposite $m$; and the increasing proton-proton entanglement with neutron excess, are  even more evident in neon isotopes. Figure~\ref{fig:SijNe} shows the mutual information for  $^{20-28}$Ne, using the same structure explained in the first panel of Fig.~\ref{fig:SijBe}. The values of $S_{ij}$ for each isotope can again be roughly understood in terms of a naive filling of the three subshells in the $sd$ shell, $0d_{5/2}$, $1s_{1/2}$, and $0d_{3/2}$, with 6, 2 and 4 single-particle orbitals, respectively. For ${}^{20}$Ne (leftmost panel), the largest neutron-neutron correlations appear in the lowest subshell, while for ${}^{28}$Ne (rightmost panel), with the two lowest subshells mostly full, the largest neutron-neutron mutual information is among the $0d_{3/2}$ states. Just as in beryllium, proton-proton entanglement in neon also increases notably with neutron number. Indeed, Figure~\ref{fig:SijNe} shows that the bottom-left blocks, corresponding to proton $0d_{5/2}$ states, become darker as the number of valence neutrons increases. Similarly to what was observed in beryllium isotopes, proton-neutron correlations in neon are almost negligible in comparison with like-particle correlations. Within each subshell, neon isotopes present the largest correlation among orbitals with opposite $m$, for both the proton-proton and neutron-neutron sectors. The only exception is the $0d_{3/2}$ neutron subshell in ${}^{28}$Ne (top right panel), where all orbitals present relatively similar and large entanglement. Our mutual information for $^{26}$Ne is in very good agreement with the ab initio valence space results presented by Tichai et al~\cite{tichai2022combining}. Likewise, our mutual information for $^{24}$O ---discussed below--- is also similar to the results shown in Ref.~\cite{tichai2022combining}.

Nonetheless, Fig.~\ref{fig:SijNe} indicates that the mutual information between proton modes in $^{20}$Ne and $^{22}$Ne is qualitatively different: here opposite $m$ modes do not dominate as much as for heavier isotopes. This suggest that pairing correlations are not that dominant for these deformed nuclei, as they are governed by quadrupole correlations~\cite{Elliott1,Elliott2}. This dominance is highlighted by the similar mutual information in the $d_{5/2}-s_{1/2}$ subshells, orbitals with $\Delta l=2 $ which accommodate most of the correlations~\cite{Zuker1995}. Figure~\ref{fig:SijBe} shows a similar picture with quite comparable and relatively small mutual information for several modes for $^8$Be, which is also a deformed nucleus. Hence, the entanglement structure given by the mutual information measure can help us discern collective deformation effects even in relatively small configuration spaces.
\begin{figure*}[t]
     \centering
     \includegraphics[width=\linewidth]{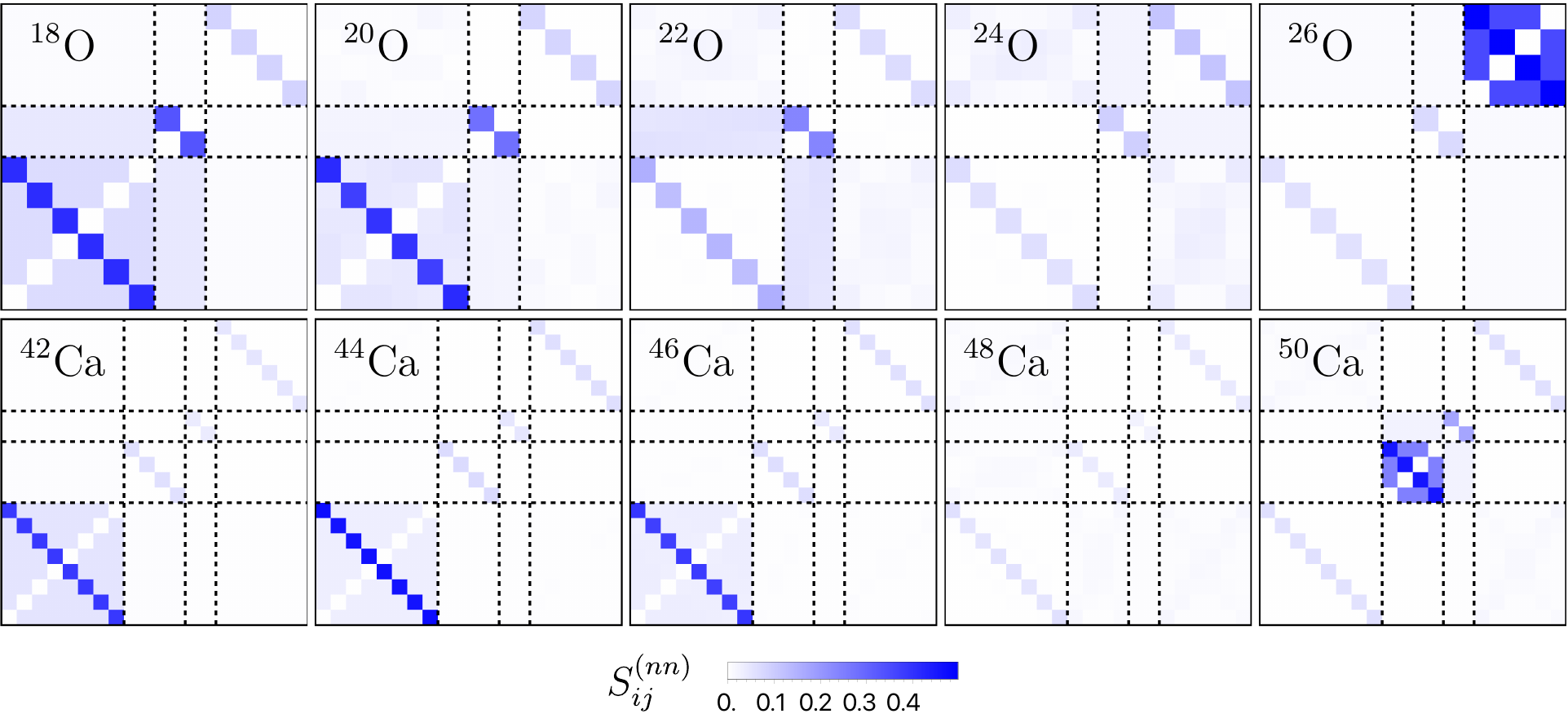}
     \caption{Mutual information $S_{ij}$ for $^{18-26}$O (top) and $^{42-50}$Ca (bottom). 
     The orbital numbering follows the scheme of Fig.~\ref{fig:SijBe} for single-neutron orbitals. Dashed lines correspond to the subshells of the $sd$ shell for oxygen, as indicated in Fig.~\ref{fig:SijNe}, and to the $0f_{7/2}$, $1p_{3/2}$, $1p_{1/2}$ and $0f_{5/2}$ subshells of the $pf$ shell, from left to right (and bottom to top), for calcium.
    }
     \label{fig:SijOCa}     
\end{figure*}

We continue our analysis by focusing on two additional isotopic chains. 
Figure~\ref{fig:SijOCa} shows the mutual information
for $^{18-26}$O (top row) and $^{42-50}$Ca (bottom row). These nuclei contain only valence neutrons in the $sd$ and $pf$ shells, respectively.
In these cases, the entanglement of the opposite-$m$ partitions  is even more clear than for beryllium and neon, as shown by the strong diagonals appearing in each subshell block. These diagonals clearly stand out above the rest of the correlations.

As discussed earlier, the entanglement in each subshell depends strongly on the
number of valence neutrons.
For the lightest oxygen isotopes, ${}^{18}$O and ${}^{20}$O, with 2 and 4 valence neutrons,
the 6 orbitals in the
$0d_{5/2}$ shell are roughly half filled and present large orbital-orbital entanglement. 
Likewise, the  $0f_{7/2}$ orbitals in  
${}^{42}$Ca, ${}^{44}$Ca, and ${}^{46}$Ca, show substantial mutual information. 

In contrast, ${}^{24}$O has the $0d_{5/2}$ and $1s_{1/2}$ subshells mostly filled, and the remaining valence orbitals are mostly empty. Consequently, the mutual information across all orbitals is small. 
Equivalently, ${}^{48}$Ca, with 8 valence neutrons, has a mostly full $0f_{7/2}$ subshell. ${}^{24}$O and ${}^{48}$Ca thus present
low mutual information in all subshells, as expected from the single-orbital entanglement values $S_i$ in Table~\ref{tab:Si}. In addition, we observe that ${}^{48}$Ca shows the lowest mutual information computed in this work, in agreement with the nuclear-structure viewpoint as this nucleus is double magic. 

The heaviest isotope studied in these two isotopic chains,
${}^{26}$O and ${}^{50}$Ca, correspond again
to half-full subshells in a naive shell model ordering.
For ${}^{26}$O, this is the $0d_{3/2}$ subshell (top-right block), whereas for ${}^{50}$Ca  
it is the $1p_{3/2}$ subshell (second antidiagonal block).

Similarly to ${}^{28}$Ne, these two isotopes present large mutual information across the whole half-filled $j=3/2$ subshell (see the darker blocks in the two rightmost panels of Fig.~\ref{fig:SijOCa}). 

Overall, the mutual information shows several important features.
First, entanglement is largest between orbitals with opposite $m$~\cite{Legeza:2015fja,tichai2022combining}.
This is to be expected from nucleon-nucleon pairing correlations, as the interaction enhances the formation of isovector nucleon pairs which are coupled to total $J_{12}=0$, or equivalently, $m_1+m_2=0$~\cite{dean2003pairing,brink2023nuclear}. Previous studies including pairing correlations in quantum simulations have been performed on the Agassi model~\cite{illa2023quantum,perez2022digital,saiz2022digital}. In contrast, in deformed nuclei the mutual information is qualitatively different, being relatively similar between different modes and typically small.

Further, the entanglement between proton and neutron orbitals is notably low in comparison with like-particle orbitals, as previously observed in Ref.~\cite{johnson2023proton}.
Furthermore, as the number of excess neutrons increases, proton-neutron entanglement
diminishes while protons become more entangled among themselves.
We only observe subtle hints of proton-neutron entanglement in cases with nearly the same number of protons and neutrons~\cite{frauendorf2014overview,romero2019symmetry,lei2011systematic}, such as $^8$Be or $^{20}$Ne.

\subsection{Equipartition entanglement}

The mutual information studied in section~\ref{sec:mutual} provides a global picture of the entanglement structure of different  nuclei. 
This analysis, however, is restricted to local, orbital-orbital correlations.
To understand whether these features translate into low or high entanglement among
all proton and neutron orbitals ($S_{pn}$), or among all $m<0$ with all $m>0$ modes ($S_m$), we additionally compute
the von Neumann entropies for these two specific equipartitions.

Table~\ref{tab:entropies} collects the values of $S_{pn}$ and $S_m$ for all nuclei studied in this work.
We compare these entanglement measures to their potential maximum values determined by the Fock subspace, as discussed in Sec.~\ref{sec:max_entropy}, and show the relative entropies $S_{pn}/S_{pn}^{(max)}$ and $S_{m}/S_{m}^{(max)}$
 in parenthesis.
We find $S_{pn}<2$ for all beryllium and neon isotopes, 
which corresponds to less than half of the maximum bound for $S_{pn}$.
The entanglement between all proton and neutron orbitals is indeed low, both in absolute and relative value, compared to the corresponding maximum. Importantly, this proton-neutron entanglement measure decreases with neutron excess.

In contrast, the values of  $S_m$ are relatively
large for all nuclei. In particular, for light nuclei, $S_m$ is close to saturating the bound. 
This is to be expected from the mutual information values of Figs.~\ref{fig:SijBe},
\ref{fig:SijNe}, and \ref{fig:SijOCa}. The isotopic dependence of $S_m$ is richer than that of $S_{pn}$. In particular, it reflects the corresponding subshell closures of isotopes like ${}^{22}$O, ${}^{24}$O and ${}^{48}$Ca.
This indicates that the energy and spin structure of the valence shell has a larger influence on  the amount of entanglement $S_m$ than what would be expected from $S_{max}$, the maximum potential entanglement given a number of orbitals and nucleons.
The latter is always maximal, when computed as explained in Sec.~\ref{sec:max_entropy},
when the shell is half full, while $S_m$ is largest when particular $subshells$ are half full.

\begin{table}[t]
\begin{center}
\begin{tabular}{c|c|c|c|c}
   \text{Nucleus} & $S_{pn}$ & $S_m$ & $\overline{S}$ & $\sigma_S$
  \\ \hline
  $^8{}$Be    &  1.99 (0.51) &  3.67 (0.91) & 3.84 & 0.22  \\
  $^{10}{}$Be &  1.05 (0.27) &  3.04 (0.94) & 3.04 & 0.27 \\
  $^{12}{}$Be &  -           &  1.42 (1.00) & 1.32 & 0.24 \\\hline
  $^{18}{}$O  &  -           &  2.22 (0.86) & 1.99 & 0.31 \\
  $^{20}{}$O  &  -           &  2.85 (0.67) & 2.66 & 0.34 \\
  $^{22}{}$O  &  -           &  1.67 (0.35) & 1.66 & 0.15 \\
  $^{24}{}$O  &  -           &  0.80 (0.19) & 0.80 & 0.08 \\
  $^{26}{}$O  & -            &  1.41 (0.55) & 1.30 & 0.24 \\ \hline
  $^{20}{}$Ne & 0.77 (0.13)  &  5.08 (0.81) & 5.15 & 0.18 \\
  $^{22}{}$Ne & 1.25 (0.21)  &  5.80 (0.76) & 5.88 & 0.28 \\
  $^{24}{}$Ne & 1.27 (0.21)  &  4.86 (0.61) & 4.86 & 0.29 \\
  $^{26}{}$Ne & 0.51 (0.08)  &  4.20 (0.57) & 3.99 & 0.31 \\
  $^{28}{}$Ne & 0.27 (0.04)  &  3.70 (0.63) & 3.51 & 0.35 \\ \hline
  $^{42}{}$Ca & -            &  2.43 (0.73) & 2.04 & 0.35 \\
  $^{44}{}$Ca & -            &  3.38 (0.58) & 2.87 & 0.47  \\
  $^{46}{}$Ca & -            &  3.00 (0.40) & 2.62 & 0.37  \\
  $^{48}{}$Ca & -            &  0.96 (0.11) & 0.91 & 0.06  \\
  $^{50}{}$Ca & -            &  2.39 (0.27) & 2.15 & 0.29 
\end{tabular}
\end{center}
\caption{Von Neumann entanglement entropies for the proton-neutron, $S_{pn}$, and opposite $m$, $S_m$, partitions
(second and third columns). The numbers in parenthesis are the entropies normalized to the maximum possible value
constrained by the Fock subspace.
The fourth and fifth columns show the average entropy and standard deviation 
of all the calculated equipartitions. In the case of neon isotopes, we have taken a sample of 1\% of all possible equipartitions.
}
\label{tab:entropies}
\end{table}

It is also interesting to quantify how the entanglement of the proton-neutron and $S_m$ partitions compares to the entanglement of all the other partitions.
To this end, we compute the entropies for all possible equipartitions for ${}^{8}$Be and ${}^{10}$Be,
consisting of 12 single-particle orbitals in the $p$ shell.
This implies a total of 
$\frac12\binom{12}{6}=462$ 
equipartitions for these isotopes.

\begin{figure}[t]
    \centering
    \includegraphics[width=1.\linewidth]{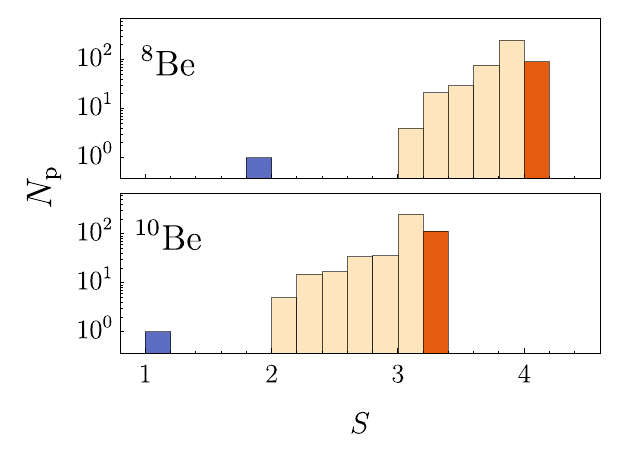}
    \caption{Distribution of von Neumann entanglement entropies for all possible
    equipartitions of $^{8}$Be (top) and $^{10}$Be (bottom) in the $p$ shell. The separated single-count blue bars in each panel correspond to proton-neutron partitions. In both histograms, the opposite $m$ partition falls within the red (darker) bars.} 
    \label{fig:histBe}
\end{figure}

Figure~\ref{fig:histBe} shows a histogram representing the distribution of all the von Neumann entropies associated to all these partitions. 
There are several remarkable properties in this plot that happen to be relatively robust across all the other isotopic chains. 
The equipartition histogram is asymmetric, akin to a  skewed normal distribution, with a sharp decay past the maximum. 
We show the bin corresponding to $S_{pn}$ in a different colour (blue), to highlight the fact that this is the lowest
of all possible equipartition entropies in both nuclei.

We also emphasize the partition of $m<0$ and $m>0$ orbitals, using a red histogram bar in the two panels of Fig.~\ref{fig:histBe}. 
For $^8$Be, as discussed in Sec.~\ref{sec:max_entropy}, the maximum possible entropy is $S \approx 3.9$.
The von Neumann entropy for the opposite $m$ partition falls, for the two isotopes, in the bar at the very right of the histogram. This indicates that the opposite $m$ partition presents
almost maximal entanglement.

Finally, we find a significant isotopic dependence on the von Neumann entropy distribution of equipartitions. We find a general shift when going from
${}^{8}$Be (top panel)
to ${}^{10}$Be (bottom panel). 
We note that this difference is unique to
equipartition entanglement. It is not, for instance,
observed in the mutual information plots of Fig.~\ref{fig:SijBe},
where $S_{ij}$ is larger for ${}^{10}$Be than for ${}^{8}$Be.
In fact, if we compute the average values of $S_{ij}$,
with $i\neq j$, we obtain $\langle S\rangle_{ij}=0.029$  for ${}^{8}$Be and  $\langle S\rangle_{ij}=0.043$ for ${}^{10}$Be.
These are in contrast to the mean entropies obtained from the average of the data in Fig.~\ref{fig:SijBe}. These are reported
in column 4 of Table~\ref{tab:entropies}. 
We indeed find that the average entropy decreases 
from a value of $\overline{S}=3.84$ in  ${}^{8}$Be 
to $\overline{S}=3.04$ in ${}^{10}$Be. 
We conclude that 
a nucleus can have more entanglement localized in specific orbitals than another one, and yet have an overall smaller multi-orbital entanglement. 

\begin{figure}[t]
    \centering
    \includegraphics[width=1.\linewidth]{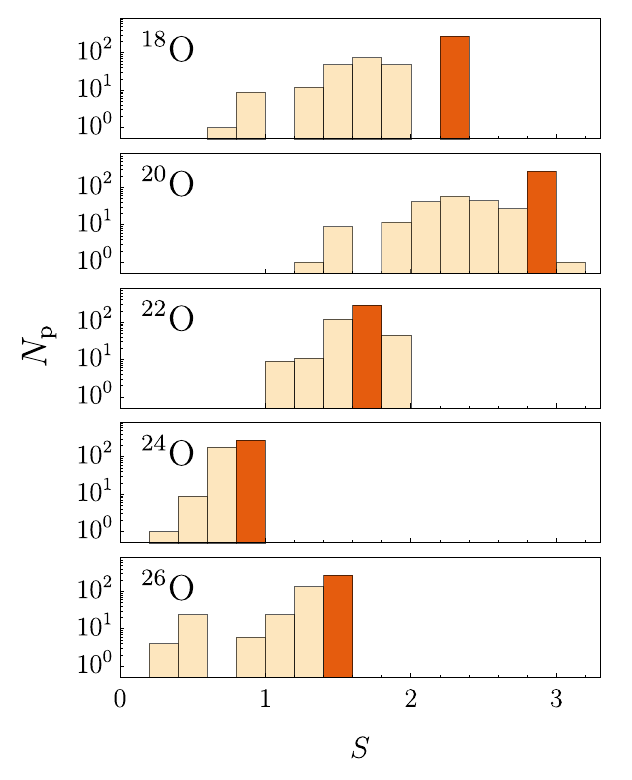}
    \caption{Same as in Fig.~\ref{fig:histBe}, for $^{18-26}$O.
    } 
    \label{fig:histO}
\end{figure}

Figure~\ref{fig:histO} shows the corresponding equipartition entropy distributions for the oxygen isotopic chain, from $^{18}$O to $^{26}$O.
In this case the valence spaces consist of only neutron orbitals, so there is no proton-neutron partition.
There is a total of $\frac12\binom{12}{6}=462$  available equipartitions. 
The oxygen entropy distributions present more structure than those of beryllium. 
For ${}^{18}$O (top panel), the distribution has some gaps, and the largest entropy bin is also the most populated.
${}^{20}$O has the largest equipartition entanglement, as measured by the mean value reported it Table~\ref{tab:entropies}. It also has the broadest distribution, as quantified by means of the standard deviation, shown in column 5 of Table~\ref{tab:entropies}. The largest standard deviation across the oxygen isotopic chain is indeed the one associated with ${}^{20}$O.

As the neutron number increases past ${}^{20}$O, the distribution changes in shape and structure. The mean entropy decreases for ${}^{22}$O and is at its lowest in ${}^{24}$O. This is consistent with the top panels of Fig.~\ref{fig:SijOCa},
and expected from the $0d_{5/2}$ and $1s_{1/2}$ subshell closures.
Moreover, the distributions for these isotopes have a significantly lower standard deviation.
Beyond the subshell closure, ${}^{26}$O shows a broader distribution, with a two-peak structure and an overall larger mean. 
In oxygen, the particular equipartitions  corresponding to $m<0$ and $m>0$ orbitals show, again, an almost maximal von Neumann entropy. This can be clearly identified in the plots, where the red bars tend to appear at the right of the histograms.

\begin{figure}[t]
    \centering
    \includegraphics[width=1.\linewidth]{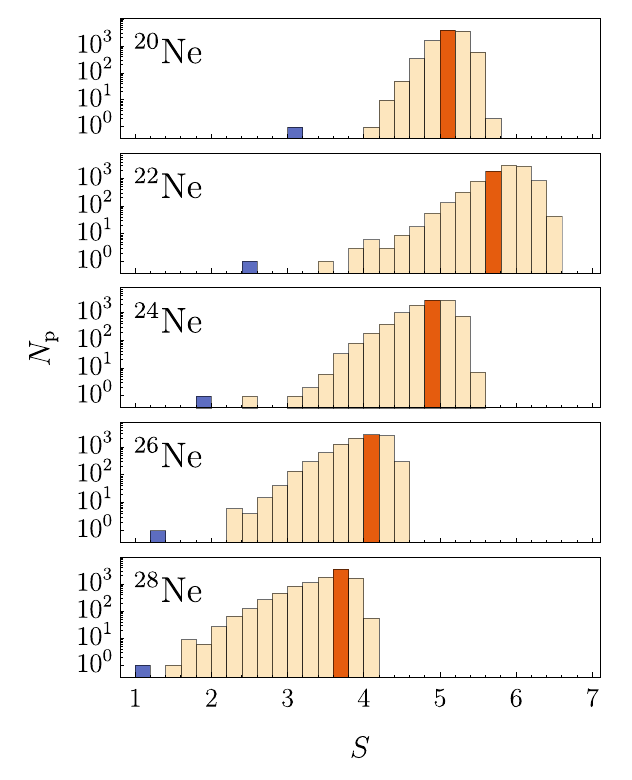}
    \caption{
    Same as in Fig.~\ref{fig:histBe}, for $^{20-28}$Ne. Here $1\%$ samples of all equipartitions have been calculated.
    }
    \label{fig:histNe}
\end{figure}

To study whether the proton-neutron partition provides the lowest entanglement
among all equipartitions in the $sd$ shell, we show in Fig.~\ref{fig:histNe} the von Neumann entropy distributions for $^{20-28}$Ne.
Neon isotopes have a total of 24 orbitals, resulting in a total of
$\frac12\binom{24}{12}\simeq1.4\cdot 10^6$ 
equipartitions. 
In general, the number of equipartitions scales combinatorially as $\frac12\binom{N_q}{N_q/2}$, where $N_q$ is the number of qubits, corresponding to orbitals in the configuration space under the Jordan-Wigner mapping. The scaling prevents the computation of the entropy for all possible equipartitions already for neon isotopes in the \emph{sd} shell. Once all equipartitions have been defined, which is relatively fast,
we take a random sample of 1\% of all these equipartitions to generate the results in the figure, using an uniform probability distribution.
As in Fig.~\ref{fig:histBe}, the proton-neutron
entropy, marked in blue,
appears well separated from the rest of the distribution in all neon isotopes.
This highlights again the uniqueness of this partition. 

The overall shape of the distribution for neon isotopes is more reminiscent of a Gaussian, although with a sharp cutoff at high entropies. 
The mean and the standard deviation of the distribution increases when going from ${}^{20}$Ne to ${}^{22}$Ne, just as it did with the oxygen isotopes (see Table~\ref{tab:entropies}). Beyond this point, the mean of the distribution steadily decreases with neutron number, even past the $0d_{5/2}$ subshell closure. 
In fact, the distributions for ${}^{20}$Ne (top panel) and ${}^{28}$Ne (bottom panel) barely overlap. 

In each of these isotopes, the equipartition between opposite $m$ orbitals belongs to a histogram (marked in red) that
falls roughly at the peak of the distribution and follows the corresponding means, sitting just below only for deformed $^{20-22}$Ne, see also Table~\ref{tab:entropies}.
This is in stark contrast to the entropy of this very same partition for beryllium and oxygen, where the same equipartition sat close to or at the bin with highest entropy.
This difference can be understood assuming that most of the equipartition entanglement is local.
Mutual information figures indicate that, 
if all entanglement were local,
all the equipartitions which split the same set of
pairs with same energy and opposite $m$, independently of which part of the pair falls in which equipartition,
would have the same equipartition entanglement.
Entanglement of equipartitions which split different number of $m$-pairs would vary due to the the differences in entanglement of the included or excluded pairs.
The $S_m$ partition in particular is one out of many ways
to split all the possible $m$-pairs into two equipartitions. Moreover, this number grows with the number of orbitals in the valence shell.
In this simplified picture, it is expected that a broader distribution surrounds $S_m$ for neon isotopes,
with 24 orbitals, than for beryllium and oxygen isotopes, with 12 orbitals.

For brevity purposes, we do not report on the distributions of calcium isotopes. These follow a similar shape than those of oxygen isotopes. They also present gaps in the spectrum,
and a shift in their mean entropy in correspondence with the bottom panels of Fig.~\ref{fig:SijOCa}. In this case, the maximum mean entropy peaks at ${}^{44}$Ca, with $\overline{S}=2.87$. This isotope also presents the broadest distribution, with $\sigma_s=0.47$. On the ohter hand, ${}^{48}$Ca has the lowest mean entropy and standard deviation, with
$\overline{S}=0.91$ and 
$\sigma_s=0.06$.

Let us finally discuss the overall behaviour of the means and standard deviations of these von Neumman entropy
distributions.
These are presented in the fourth and fifth
columns of Table~\ref{tab:entropies}, including also values for the calcium isotopic chain.
These statistical metrics provide a quantitative measure of non-local entanglement for each nucleus.
In general, we find that these numbers correlate with the results we discuss in previous subsections, and with general nuclear structure wisdom. 
Beryllium isotopes, in the $p$ shell, show a decrease of von Neumann mean entropy and standard deviation with neutron excess. 
For semimagic oxygen and calcium, with closed-shell protons, the mean von Neumann equipartition entropy is largest in mid-subshell isotopes like $^{20}$O and $^{44}$Ca. As the neutron number increases past this point, the von Neumann entropy decreases as it reaches the corresponding subshell closure isotopes, $^{22}$O, $^{24}$O and $^{48}$Ca, being minimum in the last two nuclei. This is not only true for the central values, but also for the standard deviations, which peak around the midshell maximum and are the smallest in the corresponding subshell closures. 
These subshell structures are more difficult to ascertain in the neon isotopic chain. This is naively expected from a nuclear structure point of view, since correlations smear the corresponding neutron and proton single-particle structures.

\section{Conclusions}
\label{sec:conclusions}

In this work, we analyze entanglement features in the nuclear shell model, with focus on Be, O, Ne and Ca isotopes. We use different metrics to quantify the importance of entanglement, including single-orbital entropies, orbital-orbital mutual information, and the von Neumann entropies between two equipartitions of the valence space.
In all cases, we find that the entanglement properties are sensitive to the nuclear structure and depend, in some cases strongly, on the (valence) neutron and proton numbers. 
Nonetheless, different entanglement metrics reflect different correlation features within the system.

Single-orbital entanglement depends
strongly on the energy, angular momentum, and isospin of the corresponding orbitals.
It is mainly a reflection of the evolution of the single-particle occupation numbers, which is relatively well understood based on nuclear structure insights. When, on the other hand, a nucleus shows single-orbital entanglements in contrast with the naive filling of single-particle orbitals, this can be an indication of deformation.

Orbitals with either very small or very large occupation numbers, however, can only have a limited contribution to many-orbital entanglement, as computed
with the mutual information or equipartition entropies. This is consistent with the discussion in Sec.~\ref{sec:max_entropy} on how the allowed many-body states limit the construction of states following Eq.~(\ref{eq:max_ent}). 
In general, we find that mutual information gives a good overall picture of the entanglement
structure. Mutual information displays various key explicit features across the $p$, $sd$ and $pf$ shells. First, there is an extremely low proton-neutron entanglement, compared to like-particle entanglement. Second, in spherical nuclei the proton-proton and neutron-neutron pairs with the largest mutual information are those with the same single-particle energy, but opposite third-component of the total angular momentum, $m$. Finally, deformed nuclei are characterized by rather constant mutual information between modes, with relatively low values.

These features are not unique to the mutual information metric, but turn out to be relatively generic. We see these reflected, for instance, in the distribution of von Neumann entropies corresponding to all the possible equipartitions in the system. In all cases studied so far, we find that the proton-neutron partition presents the lowest entanglement. 
Moreover, we find that, for all available measures, the proton-neutron entanglement decreases with neutron excess.
This indicates that, in order to simulate separately two halves of the valence space, the optimal choice 
is to split this space in terms of the isospin projection, $t_z$.
This is in agreement with and extends previous findings~\cite{johnson2023proton}.
Opposite $m$ partitions, in contrast, are close to the maximum allowed entropies. For most of the isotopes studied here, we find that the opposite $m$ partition is more than $50 \%$ of the maximum bound imposed by the dimension of the Fock space. 

These results showcase future possible avenues of work.
First, on the nuclear structure side, these very same techniques could be employed for odd nuclei, whose nuclear structure is not so much driven by nuclear pairing compared to even-even systems. We also plan to perform a closer analysis of the entanglement signatures of nuclear deformation. It would also be interesting to analyze the nuclear structure of the same nuclei studied in this work within the no-core shell model,
testing if, and how, entanglement measures can identify the appearance of a core and a valence space.

 Second, on the entanglement quantification front, 
one may use other entanglement measures, like $n-$tangles, to give a further insight into the topic. This is particularly relevant in relation to multipartite entanglement in fermionic systems~\cite{momme2023}. 

Finally, it would be interesting to exploit our findings in practical circuit simulations, a task we aim to undertake in the near future. 
In particular, we plan to exploit low entanglement partitions to build independent quantum circuits that allow for accurate, yet less resource-intensive, results. 
Such concrete circuit proposals may also lead to new performance comparisons between different fermionic encodings, as some partitions may only be unambiguously possible in  specific encodings. More interestingly, they may pave the way for more efficient circuit designs to study atomic nuclei across the nuclear chart with quantum simulations.
Similarly, they can also be exploited in classical simulations that rely on entanglement as a resource~\cite{Legeza:2015fja,tichai2022combining}. For example, tensor networks simulations are only efficient when they contain tensors with small rank. Finding partitions with low entanglement is equivalent to identifying a structure that connects through such a low-rank tensor, meaning one can use these criteria to decide if a tensor network structure is appropriate for the simulated system.

\begin{acknowledgements}
This work is financially supported by 
the Ministry of Economic Affairs and Digital Transformation of the Spanish Government through the QUANTUM ENIA project call - Quantum Spain project, 
by the European Union through the Recovery, Transformation and Resilience Plan - NextGenerationEU within the framework of the Digital Spain 2026 Agenda,
by grants PID2020-118758GB-I00 and PID2020-114626GB-I00 
funded by MCIN/AEI/10.13039/5011 00011033; 
by the ``Ram\'on y Cajal" grants RYC-2017-22781 and RYC2018-026072 funded by MCIN/AEI /10.13039/50110 0011033 and FSE “El FSE invierte en tu futuro”;  
by the  “Unit of Excellence Mar\'ia de Maeztu 2020-2023” award to the Institute of Cosmos Sciences, Grant CEX2019-000918-M funded by MCIN/AEI/10.13039/501100011033; 
and by the Generalitat de Catalunya, grant 2021SGR01095.
\end{acknowledgements}

\bibliography{biblio}

\end{document}